\documentclass[aps,prl,reprint,showpacs,showkeys,times,superscriptaddress]{revtex4-1}
\usepackage{amsmath,amssymb,amsthm,times,graphics,graphicx,bm}
\usepackage{soul}
\usepackage{color}

\usepackage[colorlinks,linkcolor=red,citecolor=blue,urlcolor=blue]{hyperref}

\newcommand{\be}{\begin{equation}}
\newcommand{\ee}{\end{equation}}
\newcommand{\ben}{\begin{eqnarray}}
\newcommand{\een}{\end{eqnarray}}
\newcommand{\bes}{\begin{subequations}}
\newcommand{\ees}{\end{subequations}}
\newcommand{\bF}{\begin{figure}}
\newcommand{\eF}{\end{figure}}

\begin{document}
\title{Investigating the Effects of the Interaction Intensity in a Weak Measurement}

\author{Fabrizio Piacentini}
\affiliation{Istituto Nazionale di Ricerca Metrologica, Strada delle Cacce 91, 10135, Torino, Italy}
\author{Alessio Avella}
\affiliation{Istituto Nazionale di Ricerca Metrologica, Strada delle Cacce 91, 10135, Torino, Italy}
\author{Marco Gramegna}
\affiliation{Istituto Nazionale di Ricerca Metrologica, Strada delle Cacce 91, 10135, Torino, Italy}
\author{Rudi~Lussana}
\affiliation{Politecnico di Milano, Dipartimento di Elettronica, Informazione e Bioingegneria, Piazza Leonardo da Vinci 32, 20133 Milano, Italy}
\author{Federica Villa}
\affiliation{Politecnico di Milano, Dipartimento di Elettronica, Informazione e Bioingegneria, Piazza Leonardo da Vinci 32, 20133 Milano, Italy}
\author{Alberto Tosi}
\affiliation{Politecnico di Milano, Dipartimento di Elettronica, Informazione e Bioingegneria, Piazza Leonardo da Vinci 32, 20133 Milano, Italy}
\author{Giorgio Brida}
\affiliation{Istituto Nazionale di Ricerca Metrologica, Strada delle Cacce 91, 10135, Torino, Italy}
\author{Ivo Pietro Degiovanni}
\affiliation{Istituto Nazionale di Ricerca Metrologica, Strada delle Cacce 91, 10135, Torino, Italy}
\author{Marco Genovese}
\affiliation{Istituto Nazionale di Ricerca Metrologica, Strada delle Cacce 91, 10135, Torino, Italy}

\date{\today}
\begin{abstract}
Measurements are crucial in quantum mechanics, in fundamental research as well as in applicative fields like quantum metrology, quantum-enhanced measurements and other quantum technologies.
In the recent years, weak-interaction-based protocols like Weak Measurements and Protective Measurements have been experimentally realized, showing peculiar features leading to surprising advantages in several different applications.
In this work we analyze the validity range for such measurement protocols, that is, how the interaction strength affects the weak value extraction, by measuring different polarization weak values measured on heralded single photons.
We show that, even in the weak interaction regime, the coupling intensity limits the range of weak values achievable, putting a threshold on the signal amplification effect exploited in many weak measurement based experiments.
\end{abstract}

\maketitle

{\it Introduction.}
The fundamental role of measurement in quantum mechanics is undisputed \cite{gen}, since it's the process in which some of the distinctive traits of the quantum world with respect to the classical one appear, e.g. the fact that quantum states collapse in a specific eigenstate of the observable (corresponding to the measured eigenvalue) when a strong measurement (described by a projection operator) is performed, causing the impossibility to measure non-commuting observables on the same particle.\\
Anyway, in the recent years a new paradigm of quantum measurement emerged, in which the coupling strength between the measured quantum state and the measurement system is weak enough to prevent the wave function collapse (at the cost of extracting only a small amount of information from a single measurement).
It is the case of Weak Measurements (WMs), introduced in \cite{2,elirev} and firstly realized in \cite{3,3a,5}, and Protective Measurements (PMs), originally proposed within the debate on the reality of the wave function \cite{Prot1} and recently realized for the first time \cite{prot_np}.\\
WMs can give rise to anomalous (imaginary and/or unbounded) values, whose real part is regarded as a conditional average of the observable in the zero-disturbance limit \cite{w1}, while the imaginary one is related to the disturbance of the measuring pointer during the measurement \cite{w2}.
Beyond having inspired a significant analysis of the meaning of quantum measurement \cite{misc1,misc2,misc4,misc5,misc6,misc7,misc8,w3}, they have been used both to address foundational problems \cite{4}, like macrorealism \cite{goggin,avella} and contextuality \cite{pusey,pusey2,hase}, and as a novel, impressive tool for quantum metrology and related quantum technologies, allowing high-precision measurements (at least in presence of specific noises \cite{4f,hall}), as the tiny spin Hall effect \cite{5} or small beam deflections \cite{6,6a,6b,6c}, and characterization of quantum states \cite{7a,7b}.
Furthermore, the absence of wave function collapse in WMs allows performing sequential measurements of even non-commuting observables on the same particle \cite{seq,piace1,thekka,misc3}, a task forbidden within the strong measurement framework in quantum mechanics.\\
On the other hand, PMs combine the weak interaction typical of WMs with some protection mechanism preserving the initial state from decoherence. Although a very controversial and debated topic from the foundational perspective \cite{Rovelli,Unruh,d,Meaning,Dass,Uffink,Prot4,Traj2,Diosi,Sch,PPP}, PMs have demonstrated unprecedented measurement capability, allowing to extract the quantum expectation value of an observable in a single measurement on a single (protected) particle \cite{prot_np}, a task usually forbidden in quantum mechanics.\\
Both of these protocols are based on a von Neumann interaction, characterized by a very weak coupling, between the observable that one wants to measure and a pointer observable.
Anyway, the regime in which the weak interaction approximation can be considered valid has not been investigated yet.
This is of the utmost relevance specially when dealing with WMs giving anomalous values, for which the weakness of the von Neumann interaction becomes crucial for the reliability of the measurement, giving rise to a signal amplification effect already demonstrated in several experiments \cite{4f,5,hall,6,6a,6b,6c}.\\
The purpose of this work is to investigate the response of the weak value measurement process in different conditions, observing, for a given interaction strength, the limits in which the expected weak value can be accurately extracted.\\

{\it Theoretical framework.}
The weak value of an observable $\widehat{A}$ is defined as $ \langle \widehat{A}\rangle_w = { \langle \psi_f | \widehat{A} |\psi_i \rangle \over \langle \psi_f | \psi_i \rangle}$, where $ | \psi_i \rangle$ and $|\psi_f \rangle$ are the pre- and post-selected quantum states, respectively.
To extract the weak value, one usually implements a von Neumann indirect measurement coupling the observable of interest (OoI) $\widehat{A}$ to a pointer observable $\widehat{P}$ by means of the unitary operation $\widehat{U}=e^{-ig\widehat{A}\otimes\widehat{P}}$, being $g$ the von Neumann coupling strength.
After a post-selection onto the state $|\psi_f\rangle$, realized by the projector $\widehat{\Pi}_f=|\psi_f\rangle\langle\psi_f|$, the information on the OoI is obtained by measuring the meter observable $\widehat{Q}$, canonically conjugated with the pointer $\widehat{P}$.\\
Considering the initial state $|\Psi_i\rangle=|\psi_i\rangle\otimes|\phi(q)\rangle$, after the von Neumann interaction and the subsequent post-selection the final state is:
\be\label{psi_fin}
|\Psi_f\rangle=\widehat{\Pi}_f\widehat{U}|\Psi_i\rangle=z\left[\mathbb{I}+ \langle\widehat{A}\rangle_w\left(e^{-ig\widehat{P}}- \mathbb{I}\right)\right]|\psi_i\rangle\otimes|\phi(q)\rangle
\ee
being $z=\langle\psi_f|\psi_i\rangle$ the internal product between the pre- and post-selected state.\\
Then, considering as initial condition $\langle\phi(q)|\widehat{Q}|\phi(q)\rangle=0$, the expectation value of the meter observable $\widehat{Q}$ onto the final state can be written as:
\begin{equation}
\label{hokuto}
\langle\Psi_f|\widehat{Q}|\Psi_f\rangle=|z|^2\Big\{2Re\left[\langle\widehat{A}\rangle_w \langle\phi(q)|\widehat{Q}e^{-ig\widehat{P}}|\phi(q)\rangle\right] + $$$$ + \left|\langle\widehat{A}\rangle_w\right|^2\left(g-2Re\left[\langle\phi(q)|\widehat{Q}e^{-ig\widehat{P}}|\phi(q)\rangle\right]\right) \Big\}
\end{equation}
In the limit of weak interaction ($g\rightarrow0$), the first perturbative order of the right term in Eq. (2) is:
\be
|z|^2g\left\{Re[\langle\widehat{A}\rangle_w] + Im[\langle\widehat{A}\rangle_w]\langle\phi(q)|\left(\widehat{Q}\widehat{P}+\widehat{P}\widehat{Q}\right)|\phi(q)\rangle
\right\}
\label{Aw}
\ee
Hence, in the case of real weak values, Eq. (2) gives:
\be
\langle\Psi_f|\widehat{Q}|\Psi_f\rangle=|z|^2g\langle\widehat{A}\rangle_w
\label{ReAw}
\ee
showing how the (real) weak value of our OoI $A$ can be obtained by a measurement of the meter $Q$, canonically conjugated with the pointer $P$.
Going further in the series expansion, one finds that the contribution at the second order is null, so the next non-trivial contribution scales as $g^3$.\\
In our experiment we extract the weak value of the polarization of single photons.
The $Im[\langle\widehat{A}\rangle_w]=0$ constraint is satisfied by restricting to pre- and post-selected states of the form $|\psi_j\rangle=\cos\theta_j|H\rangle+\sin\theta_j|V\rangle$, where $H$ ($V$) indicates the horizontal (vertical) polarization and $j=i,f$ the pre- and post-selected state.
As pointer observable we choose the transverse momentum $\widehat{P}_Q$ in the direction $Q$ (orthogonal to the photon propagation direction), then $\widehat{Q}$ will be our meter observable.
We collimate our single photons in a Gaussian mode, obtaining $|\phi(q)\rangle=\int dq f(q)|q\rangle$, with $f(q)=(2\pi\sigma^2)^{-\frac{1}{4}}\exp\left(-\frac{q^2}{4\sigma^2}\right)$.\\

\begin{figure}[t]
\begin{center}
\includegraphics[width=\columnwidth]{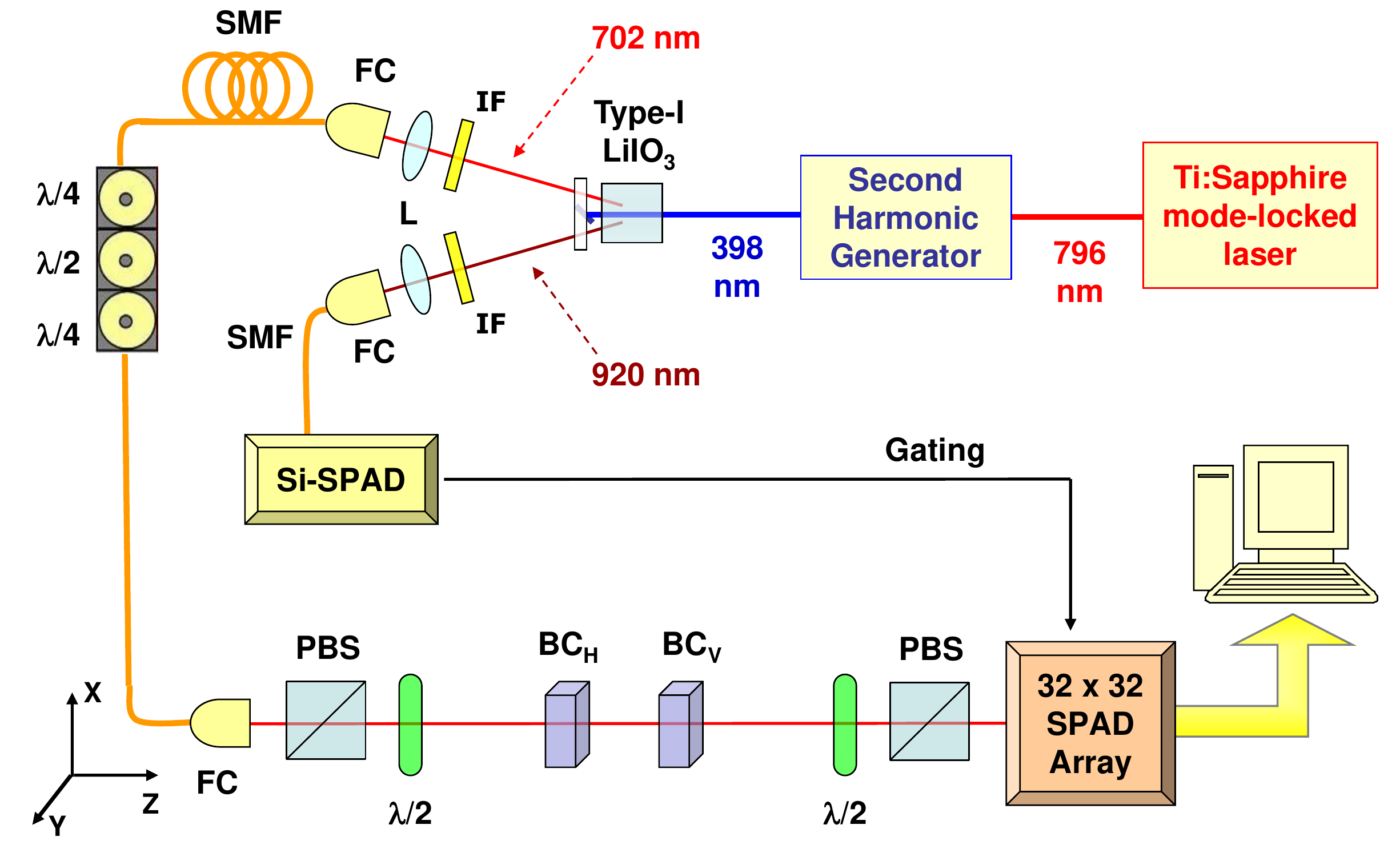}
\caption{Experimental setup. Heralded single photons are produced by downconversion in a 5-mm long type-I LiIO$_3$ non-linear crystal; the pump beam, obtained by second harmonic generation of a mode-locked laser (rep. rate 76 MHz), produces idler ($\lambda_i=920$ nm) and signal ($\lambda_s=702$ nm) photons, which pass through interference filters (IFs) before being coupled to single-mode fibres (SMFs). The idler photons are detected by means of a Silicon single-photon avalanche diode (Si-SPAD), sending a trigger pulse to the signal photons detection system. Signal photons are prepared in the initial polarisation state $| \psi_{i}\rangle$ by means of a polarising beam splitter (PBS) and a half-wave plate (HWP), then they pass through a birefringent crystal $BC_V$ shifting them in the transverse $Y$ direction, depending on their polarisation, thus measuring $\widehat{\Pi}_V$ weakly. Subsequently, an identical birefringent crystal ($BC_H$), performs the weak measurement of $\widehat{\Pi}_H$ by shifting the photons along the $X$ direction. The final post-selection onto the state $|\psi_f\rangle$ is determined by a HWP followed by a PBS. At the end of the optical path, the heralded photons are detected by a spatial-resolving $32{\times}32$ SPAD array.
}
\label{setup}
\end{center}
\end{figure}

{\it Experimental implementation.}
The single photons exploited in our experiment are produced by a heralded single-photon source \cite{IvoOE} in which a 76 Mhz Ti:Sapphire mode-locked laser at 796 nm is frequency doubled via second harmonic generation and then injected into a 5 mm thick LiIO$_3$ nonlinear crystal, generating photon pairs via type-I Parametric Down-Conversion (PDC), as reported in Fig. \ref{setup}.
The idler photon ($\lambda_i=920$ nm) is filtered by an interference filter (IF), coupled to a single mode fiber (SMF) and detected by a silicon single-photon avalanche diode (Si-SPAD).
A click from the Si-SPAD heralds the presence of the signal photon ($\lambda_s=702$ nm) in the correlated branch. Such photon passes through an IF, then is SMF coupled and addressed, collimated in a Gaussian mode, to the open air path where the weak measurements take place.
We have verified our single photon emission by measuring the antibunching parameter \cite{grangier} of our source, obtaining a value of $0.13(1)$ without any background/dark-count subtraction.\\
In such path, the heralded single photon is prepared in the linearly-polarized state $|\psi_i\rangle=\frac{1}{\sqrt2}(|H\rangle+|V\rangle)$ by means of a polarizing beam splitter (PBS) followed by a half wave plate.
After the state preparation, the photon encounters a pair of thin birefringent crystals, responsible for the weak interactions.
The first birefringent crystal (BC$_V$) presents an extraordinary ($e$) optical axis lying in the $Y$-$Z$ plane, with an angle of $\pi/4$ with respect to the $Z$ direction.
The spatial walk-off induced on the photons by BC$_V$ slightly shifts the vertically-polarized ones, separating horizontal- and vertical-polarization paths along the $Y$ direction and causing the initial state $|\psi_{i}\rangle$ to be affected by a small amount of decoherence. This element realizes the first (weak) interaction $\widehat{U}_V=e^{-ia_y\widehat{\Pi}_V\otimes{\widehat{P}_y}}$, coupling the observable under test (i.e. the vertical polarization $\widehat{\Pi}_V=|V\rangle\langle V|$) to the pointer observable, the transverse momentum along the $Y$ direction $\widehat{P}_y$.\\
Then, the photon goes through the second birefringent crystal (BC$_H$), identical to the first one, but with the $e$-axis lying in the $X$-$Z$ plane.
Here, the photons experiencing the spatial walk-off are the horizontally-polarized ones, getting shifted along the $X$ direction so that the initial polarization state undergoes the same decoherence induced by the passage in BC$_V$.
This way, the second (weak) interaction $\widehat{U}_H=e^{-ia_x\widehat{\Pi}_H\otimes\widehat{P}_x}$ occurs.
This configuration allows measuring simultaneously the weak values of the two orthogonal polarizations $\widehat{\Pi}_V$ and $\widehat{\Pi}_H$, at the same time self-compensating the unwanted temporal walk-off induced by the two interactions.\\
After the two birefringent crystals, the photon undergoes the postselection, that is, a projection onto the final state $|\psi_f\rangle$ realized by a half wave plate followed by a PBS.\\
The final photon detection is performed by a spatial resolving single-photon detector prototype, i.e. a two-dimensional array made of $32\times32$ ``smart pixels'' (each hosting a SPAD detector with dedicated front-end electronics for counting and timing single photons) operating in parallel with a global shutter readout \cite{VILLA2014}.
Each count by the Si-SPAD on the heralding arm triggers a 6 ns detection window in each pixel of the SPAD array, in order to heavily decrease the dark count rate and improve the signal-to-noise ratio.\\
We perform two different acquisitions, respectively with 1 mm and 2.5 mm thick birefringent crystals, in order to change the coupling strength of the weak interactions experienced by the single photons.
In each acquisition, we variate the postselection state and measure different weak values, observing the behaviour of the meter variables with respect to the weak values theoretically predicted.

\begin{figure}[ht]
\begin{center}
\includegraphics[width=\columnwidth]{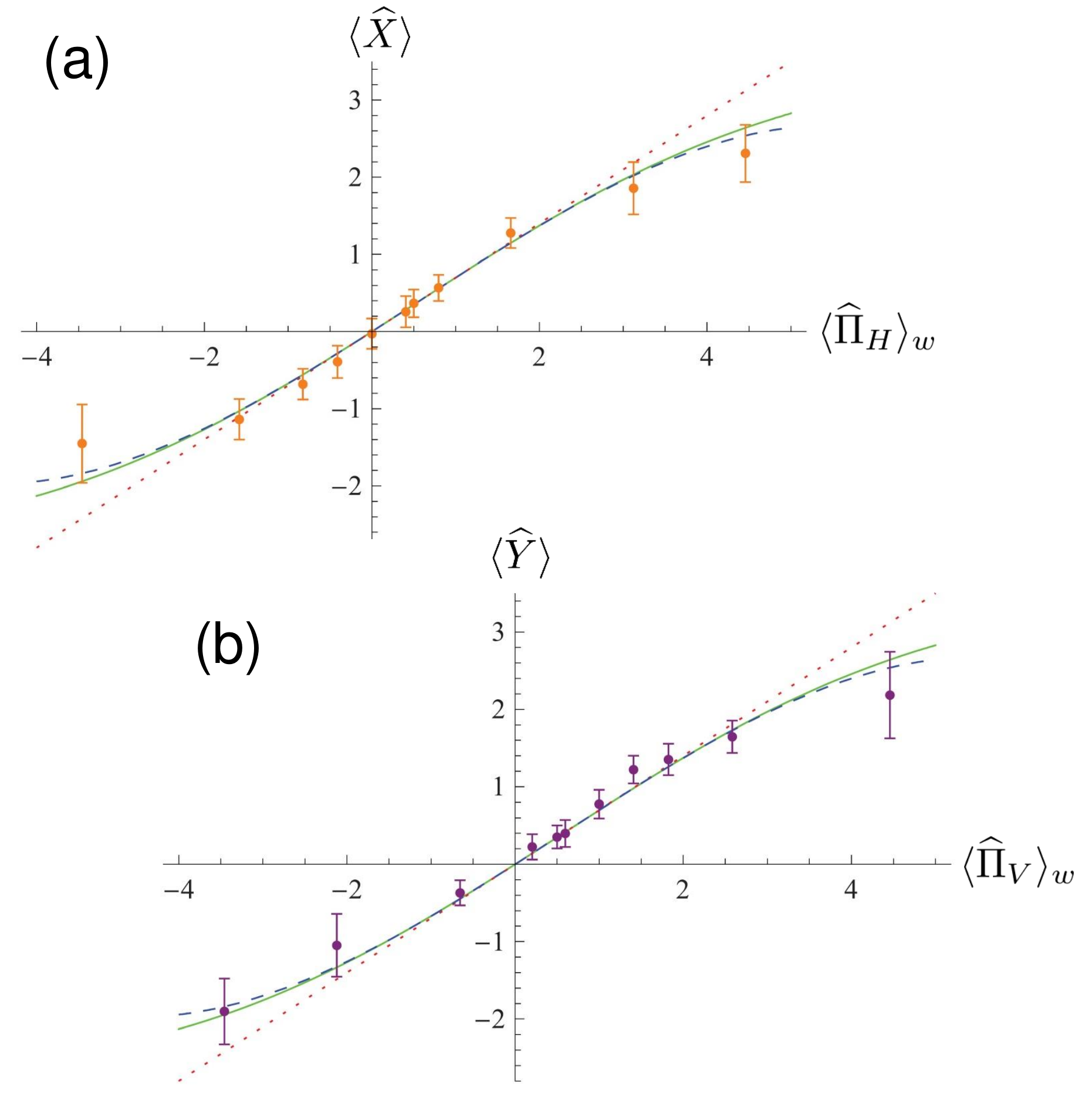}
\caption{Obtained results for the acquisition with the 1-mm long birefringent crystal pair. Plot a (b): behavior of the meter observable $\widehat{X}$ ($\widehat{Y}$) with respect to the expected weak value $\langle\widehat{\Pi}_H\rangle_w$ ($\langle\widehat{\Pi}_V\rangle_w$). Dots: experimental data. Solid green curve: complete theory of the von Neumann coupling occurring in the birefringent crystal. Dashed curve: third order approximation of the complete theory in the limit of weak coupling ($g^2\ll1$). Dotted line: first order approximation of the complete theory for $g^2\ll1$, the one used for the weak value evaluation.
}
\label{1mm}
\end{center}
\end{figure}

\begin{figure}[ht]
\begin{center}
\includegraphics[width=\columnwidth]{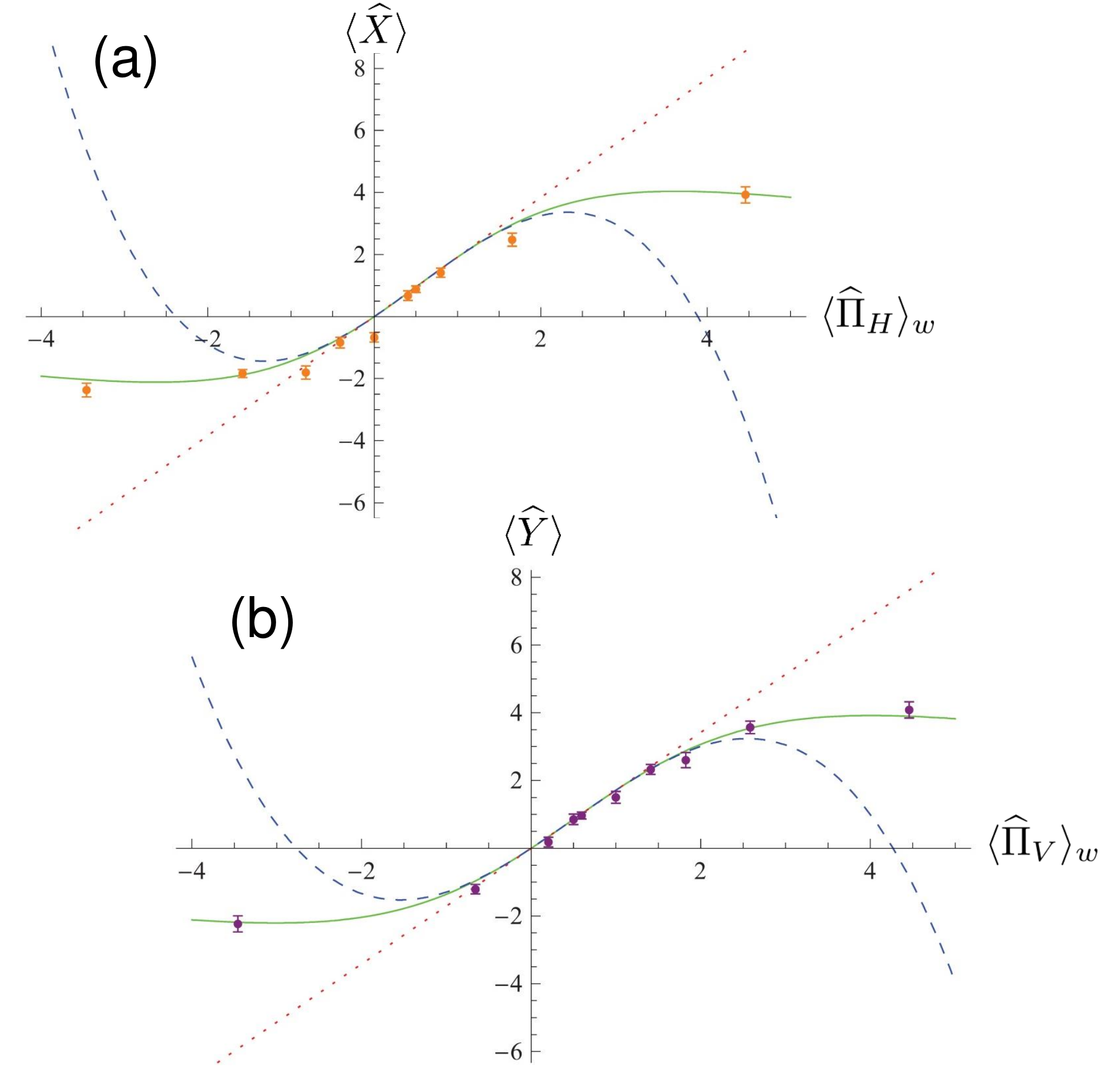}
\caption{Obtained results for the acquisition with the 2-mm long birefringent crystal pair. Plot a (b): behavior of the meter observable $\widehat{X}$ ($\widehat{Y}$) with respect to the expected weak value $\langle\widehat{\Pi}_H\rangle_w$ ($\langle\widehat{\Pi}_V\rangle_w$). Dots: experimental data. Solid green curve: complete theory of the von Neumann coupling occurring in the birefringent crystal. Dashed curve: third order approximation of the complete theory in the limit of weak coupling ($g^2\ll1$). Dotted line: first order approximation of the complete theory for $g^2\ll1$, the one used for the weak value extraction.
}
\label{2mm}
\end{center}
\end{figure}

{\it Results and conclusions.}
For each pair of birefringent crystals, we perform an initial system calibration to determine the von Neumann coupling intensity $g$, obtaining for the 1-mm long crystals $a_x=a_y=0.7$ pixels (px), while $a_x=1.9$ px and $a_y=1.7$ px for the 2.5-mm long ones (the small discrepancy between $a_x$ and $a_y$ is due to a slight mismatch in the birefringent crystals cut).
Considering that our single photons are collimated in a Gaussian distribution whose width parameter is $\sigma=4.3$ px, the two birefringent crystals pairs induce respectively an interaction strength of $g_x=g_y=a_y/\sigma\simeq0.16$ and $g_y=a_y/\sigma\simeq0.40$ and $g_x=a_x/\sigma\simeq0.45$.
These conditions should still lie within the weak interaction regime, since for all of them $g^2\ll1$.\\
The results obtained with the 1-mm and 2.5-mm birefringent crystal pairs are reported in Fig.s \ref{1mm} and \ref{2mm}, respectively.
In each of these figures, plots (a) and (b) report the behavior of the meter observables $\langle\widehat{X}\rangle$ and $\langle\widehat{Y}\rangle$ with respect to the theoretical weak values associated to them ($\langle\widehat{\Pi}_H\rangle_w$ and $\langle\widehat{\Pi}_V\rangle_w$, respectively).
The orange (purple) dots are the measured values of $\langle\widehat{X}\rangle$ ($\langle\widehat{Y}\rangle$), the solid curve represents the exact solution of Eq. (2) while the dotted line and the dashed curve indicate respectively the first order approximation, corresponding to the weak value $\langle\widehat{\Pi}_H\rangle_w$ ($\langle\widehat{\Pi}_V\rangle_w$), and the third order one in the $g^2\ll1$ limit (we remind the reader that the second order approximation gives null contribution).\\
As visible in Fig. \ref{1mm}, obtained in the condition $g\simeq0.16$ with the 1 mm birefringent crystals, the weak value approximation is valid for a good range of anomalous values, that is, for $\langle\widehat{\Pi}_H\rangle_w,\langle\widehat{\Pi}_V\rangle_w \in [-1.5,2.5]$.
Outside of this interval, instead, the data start following the third order approximation and the exact solution, almost indistinguishable in the interval investigated.
This means that, outside of the highlighted interval, a bias already affects our weak value estimation.\\
The situation becomes different when we switch to the 2.5-mm long BC$_H$ and BC$_V$, increasing the interaction strength almost to the border of the weak interaction regime.
By looking at Fig. \ref{2mm}, we can identify three regions: $\langle\widehat{\Pi}_H\rangle_w,\langle\widehat{\Pi}_V\rangle_w \in [-0.7,1.7]$ for which the meter observables still follow the weak value approximation; $\langle\widehat{\Pi}_H\rangle_w,\langle\widehat{\Pi}_V\rangle_w \in [-1.2,-0.7] \vee [1.7,2.2]$, in which the third order approximation (dashed line) is still valid; $\langle\widehat{\Pi}_H\rangle_w,\langle\widehat{\Pi}_V\rangle_w <-1.2 \vee \langle\widehat{\Pi}_H\rangle_w,\langle\widehat{\Pi}_V\rangle_w>2.2$, in which the exact solution assumes a quasi-asymptotic form and both approximations fail.
In this last region, our meter observables $\langle\widehat{X}\rangle$ and $\langle\widehat{Y}\rangle$ remain basically constant with respect to $\langle\widehat{\Pi}_H\rangle_w$ and $\langle\widehat{\Pi}_V\rangle_w$, hence it is not possible anymore to extract the weak value.\\
While in the first region one can in principle safely estimate the weak value, in the second one the bias due to the finite interaction intensity already affects such estimation, completely forbidding it in the third and last region.
This means that the signal amplification effect exploited in many WM-based experiments \cite{4f,5,hall,6,6a,6b,6c} is actually limited to a certain range of weak values, determined by the parameter to be evaluated, i.e. the interaction intensity $g$.
Outside of such interval, the weak value approximation can no longer be considered valid, forbidding any accurate weak value measurement and, as a consequence, leading to an unfaithful $g$ extraction due to biased signal amplification.\\
In the end, we experimentally investigated the limits of WMs, observing how, even in the weak interaction regime, the value of $g$ determines the range of weak values that one is able to extract, putting a threshold to the signal amplification effect \cite{4f,5,hall,6,6a,6b,6c} typical of WMs.
In fact, from our data it results evident that even a very weak coupling, satisfying the constraint $g^2\ll1$, could lead to a bias in the weak value measurement in case of strongly anomalous values.
Giving a deeper insight on weak value measurements and their properties, these results pave the way to their widespread diffusion in several applicative fields, e.g. quantum metrology, quantum-enhanced measurement and related quantum technologies.\\

\begin{acknowledgments}
This work has been supported by EMPIR-14IND05 ``MIQC2'' (the EMPIR initiative is co-funded by the EU H2020 and the EMPIR Participating States) and the MIUR Progetto Premiale 2014 ``Q-SecGroundSpace''. We thank Dr. Mattia P. Levi for contributing to the experimental setup implementation and data analysis.
\end{acknowledgments}

\end{document}